# T1-contrast Enhanced MRI Generation from Multi-parametric MRI for Glioma Patients with Latent Tumor Conditioning

Zach Eidex[1], Mojtaba Safari[1], Richard L.J. Qiu[1,3], David S. Yu[1,3],

Hui-Kuo Shu[1,3], Hui Mao[2,3] and Xiaofeng Yang[1,3*]

[1]Department of Radiation Oncology, Emory University, Atlanta, GA

[2]Department of Radiology and Imaging Sciences, Emory University, Atlanta, GA

[3]Winship Cancer Institute, Emory University, Atlanta, GA

**Running title:** T1 Contrast Synthesis

**Manuscript Type:** Original Research

**Contact information:**
Email – xiaofeng.yang@emory.edu
Address - 1365-C Clifton Road NE Atlanta, Georgia 30322

# ABSTRACT

**Objective:** Gadolinium-based contrast agents (GBCAs) are commonly used in MRI scans of patients with gliomas to enhance brain tumor characterization using T1-weighted (T1W) MRI. However, there is growing concern about GBCA toxicity. This study develops a deep-learning framework to generate T1-postcontrast (T1C) from pre-contrast multiparametric MRI.

**Approach:** We propose the tumor-aware vision transformer (TA-ViT) model that predicts high-quality T1C images. The predicted tumor region is significantly improved ($P < .001$) by conditioning the transformer layers from predicted segmentation maps through adaptive layer norm zero mechanism. The predicted segmentation maps were generated with the multi-parametric residual (MPR) ViT model and transformed into a latent space to produce compressed, feature-rich representations. The TA-ViT model was applied to T1w and T2-FLAIR to predict T1C MRI images of 501 glioma cases from an open-source dataset. Selected patients were split into training (N=400), validation (N=50), and test (N=51) sets. Model performance was evaluated with the peak-signal-to-noise ratio (PSNR), normalized cross-correlation (NCC), and normalized mean squared error (NMSE).

**Main Results:** Both qualitative and quantitative results demonstrate that the TA-ViT model performs superior against the benchmark MRP-ViT model. Our method produces synthetic T1C MRI with high soft tissue contrast and more accurately reconstructs both the tumor and whole brain volumes. The synthesized T1C images achieved remarkable improvements in both tumor and healthy tissue regions compared to the MRP-ViT model. For healthy tissue and tumor regions, the results were as follows: NMSE: $8.53 \pm 4.61E-4$; PSNR: $31.2 \pm 2.2$; NCC: $0.908 \pm .041$ and NMSE: $1.22 \pm 1.27E-4$, PSNR: $41.3 \pm 4.7$, and NCC: $0.879 \pm 0.042$, respectively.

**Significance:** The proposed method generates synthetic T1C images that closely resemble real T1C images. Future development and application of this approach may enable contrast-agent-free MRI for brain tumor patients, eliminating the risk of GBCA toxicity and simplifying the MRI scan protocol.

# 1. INTRODUCTION

Multi-parametric MRI (mp-MRI) is widely used for diagnosing and treating gliomas due to its excellent soft-tissue contrast, which, when combined with molecular and genomic biomarkers[1], enhances the characterization of these tumors. Gliomas, classified by the World Health Organization (WHO) into grades I to IV, account for over half of malignant central nervous system tumors.[2] Low-grade gliomas (LGGs; WHO grades I and II) are less aggressive and present a favorable prognosis over high-grade gliomas (HGGs; WHO grades III and IV). While non-contrast MRI sequences such as T1-weighted (T1W), T2-weighted (T2W), and T2 fluid-attenuated inversion recovery (FLAIR) provide structural information about the tumor volume and reveal structural features such as peritumoral edema, necrosis, and mass effect,[3] detailed tissue characterization remains challenging.[4] Enhancement of T1W MRI with gadolinium-based T1-contrast (T1C) improves the delineation of tumor boundaries while contrast-enhanced tumor is often indicative of HGGs. Hence, T1C is commonly incorporated into the clinical workflow.[5] However, gadolinium-based contrast agents (GBCAs) have been shown to be deposited in the brain, raising concerns about their long-term clinical implications.[6]

To address this concern associated with GBCAs and improve the availability of T1C-like images, several deep learning models have been proposed to synthesize T1C by either reducing the dose of GBCAs[7,8] or eliminating the need for contrast entirely.[9,10] However, little work has been done to improve the reconstruction of the tumor region specifically for this task. Traditional methods for enhancing tumor region reconstruction, especially for segmentation tasks, involve loss functions that assign a higher penalty to poor performance in the tumor region.[11,12] We argue that this approach is suboptimal for image translation because the model only indirectly considers information about the tumor region.

To overcome this limitation, we propose a novel approach that first predicts T1C segmentation maps from T1W and T2-FLAIR MRI using the state-of-the-art multi-parametric residual vision transformer (MPR-ViT) model[13], which leverages ViT (Vision Transformer) layers to provide global context and convolutional layers to efficiently capture finer details. Then, we train a second MRP-ViT model to generate compressed latent space representations. Finally, we introduce the tumor aware (TA) – ViT model by modifying the MRP-ViT architecture to enable the powerful attention mechanism of the ViT blocks to focus on reconstructing the tumor region. This is done by conditioning the transformer encoder on the latent segmentation maps through the adaptive layer norm (adaLN - zero) mechanism.[14]

We make the following contributions:

(1) This study is the first to use latent tumor conditioning for medical image translation, inspired by latent diffusion models[15],

(2) We show that segmentation maps can be effectively compressed into a latent space representation, reducing computational burden and eliminating superfluous information that may distract the network.

(3) The adaLN-zero mechanism efficiently guides the ViT blocks to focus on the tumor region.

(4) We achieve state-of-the-art performance in both single-modal (T1w → T1C) and multimodal (T1W + T2-FLAIR → T1C) image synthesis with TA-ViT model, noting no significant difference in reconstruction quality between the two tasks.

# 2. METHODS

## 2.1 Data Acquisition and Preprocessing

Patients were selected from the publicly available The Cancer Imaging Archive (TCIA) University of California San Francisco Preoperative Diffuse Glioma (UCSF-PDGM) dataset, containing 501 adult patients with T1w and T2-FLAIR MRI along with T1C MRI scans.[16,17] All scans were performed on a 3.0T scanner (Discovery 750, GE Healthcare, Waukesha, Wisconsin, USA) using a dedicated 8-channel head coil (Invivo, Gainesville, Florida, USA) between 2015 and 2021. The patients all had histopathologically confirmed grade II-IV diffuse gliomas: 55 (11%) grade II, 42 (9%) grade III, and 403 (80%) grade IV tumors. Segmentation maps were created from an ensemble model using top-ranking segmentation algorithms for the whole brain volume as well as for the three major tumor regions: enhancing tumor, necrotic tumor, and peritumoral edema.[18] These maps were subsequently refined by trained radiologists and verified by two expert reviewers. Bicubic interpolation was used to downsample all volumes by a factor of two to reduce the computational burden.

## 2.2 Segmentation

The T1C segmentation maps were predicted from T1W and T2-FLAIR MRI by placing them in separate input channels using the MRP-ViT architecture. While this model was orginally designed for image translation rather than segmentation, no modifications were made to the architecture or loss function as described in the original paper. The MRP-ViT model contains three parts: the encoder, the information bottleneck, and the decoder. Both the encoder and decoder are convolution-based while the information bottleneck integrates both ViT and convolutional layers. The ViT blocks employ the flash attention mechanism to improve computational efficiency.[19] All layers are connected with residual skip connections to encourage the propagation of contextually important features throughout the network [20]

## 2.3 Segmentation Map Latent Space Representation

Diffusion models have gained popularity for image translation by introducing noise into input images and then iteratively removing the noise until the final prediction is made. [21-24] This process typically involves thousands of steps and is computationally intensive, so researchers are actively exploring more efficient approaches. One promising approach is to perform the diffusion process on compressed latent representations of the images. [15] By training a network to learn the identity function, features in the information bottleneck represent compressed abstract representations of the features while removing irrelevant details in image space. These feature maps can then be used instead of the initial images in the diffusion process.

In this work, the concept of the latent space was applied to reduce the computational burden of the ViT layers, as the self-attention mechanism becomes computationally intensive depending on the number of input tokens. [25] As shown in Figure 1, the dimensionality of segmentation maps is reduced from $120 \times 120 \times 1$ to $30 \times 30 \times 256$, with the first two dimensions representing spatial dimensions and the third representing the number of channels. The MPR-ViT model was trained to learn the identity function from the ground truth segmentation maps. The ground truth latent representations were used for the training and validation datasets, while the predicted segmentation map latent representations were only used for the testing dataset.

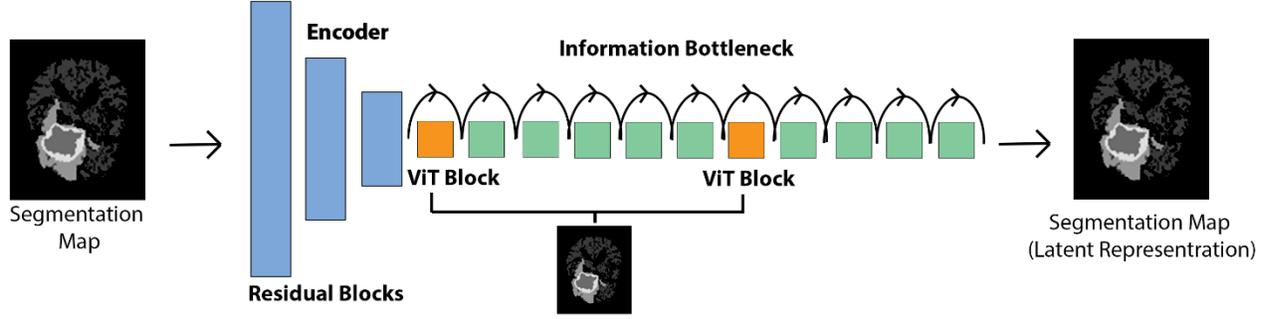

**Figure 1.** A latent representation of the segmentation maps is produced by first training the MRP-ViT model to predict the identity function. The green, blue, and orange blocks represent **two consecutive 3×3 convolutional layers, three residual blocks, and the ViT blocks, respectively.** During inference, the decoder is removed and feature maps after the last layer of the information bottleneck are gathered. The spatial dimensions of the segmentation maps are reduced from 120 ×120×1 to 30×30×256 where the first two dimensions are height and width and the last dimension is the number of channels.

## 2.4 T1C prediction with TA-ViT

Synthetic T1C MRI was predicted using the TA-ViT model. Compared to the MRP-ViT model, the primary changes are removing the convolutional layers in the ViT block so that the vision transformer architecture can work directly on the latent space representation of the segmentation map and modifying the transformer encoder to accept conditional inputs. Shown in Figure 2C, the segmentation maps along with the feature maps from the network are vectorized and given patch and positional embeddings before they are placed in the transformer encoder. The network feature maps follow a typical transformer architecture except where they are conditioned with the adaptive layer norm (adaLN).

The layer norm, defined in (1) where x is the input feature map, μ is the mean, $\sigma^2$ is the variance, $\varepsilon$ is a small constant to prevent division by zero, and $\gamma$ and $\beta$ are learnable parameters, is a standard component of the transformer architecture.[26] By summing the feature map-derived vector with the latent segmentation map vector before $\gamma$ and $\beta$ are regressed, the ViT block can be conditioned on the segmentation map or any arbitrary vector. This operation defines the adaLN block, first proposed in Peebles and Xie[14] for use in a transformer based text-conditioned diffusion model. In addition, it was found that zero-initializing $\gamma$ accelerates large scale training giving rise to the adaLN-zero block used in this work. Since transformers process 1-dimensional inputs, this mechanism could be applied to the latent segmentation maps with minimal modification and was simplified because the latent segmentation maps and feature maps are both the same dimensions. The adaLN-zero block, shown in Figure 2D was applied four times in the transformer encoder, modifying the parameters $\gamma_1$, $\beta_1$, $\alpha_1$, $\gamma_2$, $\beta_2$, and $\alpha_2$ where $\alpha$ is equivalent to the multiplicative $\gamma$.

$$LayerNorm(x) = \gamma \left( \frac{x - \mu}{\sigma^2 + \varepsilon} \right) + \beta \tag{1}$$

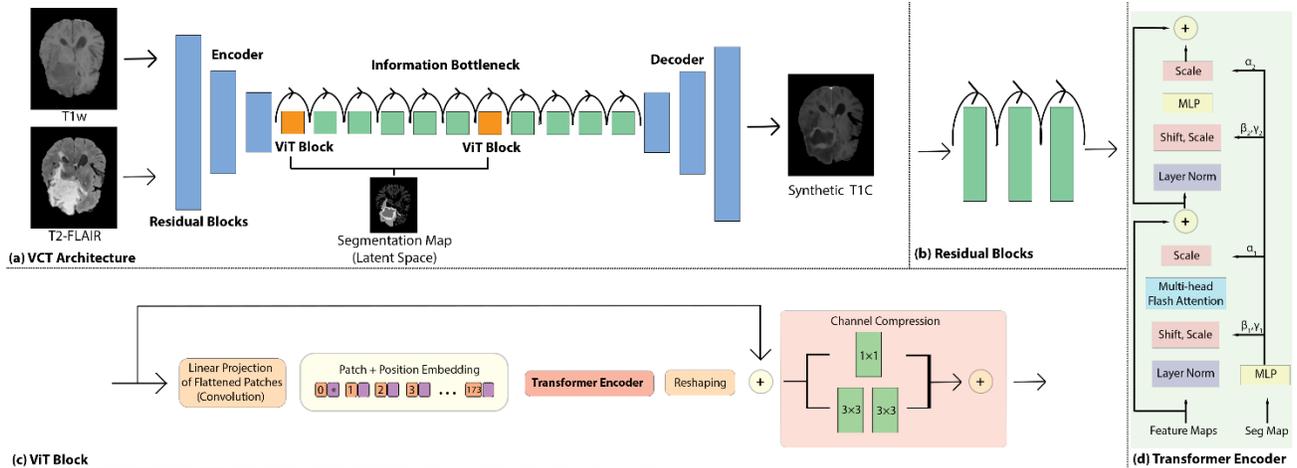

**Figure 3.** (a) Schematic flow chart of the TA-ViT model. Compared to the MPR-ViT model, the convolutional layers in the ViT Block before and after the transformer architecture were removed to retain the original latent space representation of the segmentation maps. The adaLN-zero mechanism efficiently conditions the transformer encoder to account for the tumor region by summing the latent segmentation map and feature map values before calculating the parameters α, β, and γ. The latent segmentation maps are fed into both ViT blocks.

## 2.5 Encoder and Decoder

The encoder and decoder each consist of three combined residual blocks. Within each combined residual block are three residual blocks, each comprised of a convolutional layer followed by a batch normalization layer and a Rectified Linear Unit (ReLU) activation function. All convolutional layers use a 3×3 kernel size, except for the initial and final layers which employ a larger 7×7 kernel to broaden the receptive field. Each convolutional block includes a residual skip connection to aid in gradient backpropagation.[20] In the encoder, the final convolutional layer of each combined residual block has a stride of 2, halving the dimensionality. Conversely, in the decoder, the corresponding combined residual block utilizes a transposed convolution to double the feature map size. Overall, the dimensionality is reduced by a factor of 4, bringing it down to a resolution of 30×30 before entering the information bottleneck, after which it is restored to the original input size of 120×120 within the decoder. This design strategy encourages the information bottleneck to focus on coarse details while also reducing the computational load of the ViT blocks. The encoder and decoder are designed symmetrically, with the key difference being that the final convolutional block in the decoder outputs a single channel, whereas the encoder starts with two input channels. Additionally, the network's final layer is followed by a hyperbolic tangent activation function to ensure that the output feature map values range between -1 and 1.

## 2.6 Information Bottleneck

The information bottleneck is aimed at capturing abstract, global context by using a combination of convolutional blocks and powerful, though computationally intensive, ViT blocks. To address the vanishing gradient problem, residual skip connections are employed across both types of blocks, providing an alternate path for gradients during backpropagation.

The convolutional blocks within the information bottleneck consist of two consecutive 3×3 convolutional layers, which are connected to the input feature map via a skip connection. Since the ViT layers require a 1-dimensional input, the feature maps and segmentation maps are first flattened and then individually tokenized to generate patch and positional embeddings before being fed into the transformer encoder.

As discussed in section 2.4, the adaLN-zero mechanism is used to condition the feature maps on the latent segmentation maps in the transformer encoder. The output is then connected to the input feature map through a residual skip connection.

The traditional attention mechanism has also been replaced with the more efficient FlashAttention.[19] In transformers, the main computational speed bottleneck of the attention mechanism is the read-and-write operations to GPU memory. FlashAttention addresses this challenge by optimizing the GPU's high-performance SRAM cache through tiling and recomputation, reducing memory demands while significantly boosting speed, all while maintaining computational equivalence to the traditional attention mechanism. Further efficiency gains were realized by sharing weights between the two ViT blocks.

### 2.7 Implementation Details

The MPR-ViT and TA-ViT models were trained on a consumer-grade NVIDIA RTX 4090 GPU with 24 GB of memory, and additional results were gathered with a cloud-based NVIDIA A10 with 24 GB of memory. The dataset was augmented by randomly flipping the images in the coronal plane. An AdamW gradient optimizer (learning rate $2 \times 10^{-4}$, $\beta1 = 0.500$, $\beta2 = 0.999$, eps $= 1 \times 10^{-6}$) was set to optimize the learnable parameters over 251 epochs or when the model no longer reduced the validation loss. The AdamW optimizer was chosen to minimize the loss function (L1 loss) for its improved generalization performance over the Adam optimizer due to a decoupling of the weight decay and gradient update.[27] 32 image slices were used for each batch.

### 2.8 Validation and Evaluation

Model performance was assessed using a hold-out test, where 501 patients were randomly divided into training (400 patients: 29,005 slices), validation (50 patients: 3,511 slices), and testing (51 patients: 3,590 slices) sets. The segmentation results for the tumor region were assessed using metrics including the dice similarity coefficient (DSC), the Jaccard Index (J), and the root mean squared deviation (RMSD) . Additionally, for the T1C image translation task, the evaluation metrics included the normalized mean squared error (NMSE), peak signal-to-noise ratio (PSNR), normalized cross-correlation (NCC), and structural similarity index (SSIM) over the entire 3D volume for both the entire brain and the tumor region. The student's two-sided $t$-test was employed for statistical comparison, with a significance level set at 0.05. DSC measures the overlap of the ground truth volume ($\text{VOL}_{GT}$ ) and the predicted volume ($\text{VOL}_{PT}$ ), while J measures the intersection over the union, providing a more consistent metric for tumors of different sizes. RMSD measures the average magnitude of the errors between predicted and ground truth voxel values.

Metrics to evaluate the performance of the reconstructed T1C include the NMSE, which measures the voxel-wise difference between the synthetic and ground truth volumes such that a value of zero means no difference.[28] PSNR is inversely related to the NMSE, so higher PSNR values correspond to higher similarity to the ground truth volume. Logarithmic scaling was applied to make PSNR values more closely align with human perception.[29] SSIM considers luminance, contrast, and structural similarity functions to most closely align with human perception. SSIM values range from -1 to 1 with 1 being perfect correspondence with the ground truth volume. NCC (Normalized Cross-Correlation) is a statistical measure that evaluates the correlation between the synthetic and ground truth volumes. NCC measures similarity between image structures and ranges from -1 to 1, with 1 indicating perfect correlation with the ground truth volume.[30] NMSE, PSNR, and NCC are defined below where $n$ is the total number of voxels, $X_i$ and $Y_i$ are the voxel intensity of the synthetic and ground truth volumes, and $MAX_I$ is the maximum possible voxel value of the ground truth volumes. These metrics were calculated for both the tumor region and whole brain volumes. Metrics for the tumor region were calculated by first setting all values outside of the tumor region to zero for both the ground truth and predicted volumes using the ground truth segmentation maps as the reference.

Segmentation Metrics

$$\text{DSC} = \frac{2|VOL_{GT} \cap VOL_{PT}|}{|VOL_{GT}| + |VOL_{PT}|} \tag{2}$$

$$\text{J} = \frac{|VOL_{GT} \cap VOL_{PT}|}{|VOL_{GT} \cup VOL_{PT}|} \tag{3}$$

$$\text{RMSD} = \sqrt{\frac{1}{n} \sum_{i=1}^{n}(X_i - Y_i)^2} \tag{4}$$

T1C Synthesis Metrics

$$NMSE = \frac{1}{n} \sum_{i=1}^{n}(X_i - Y_i)^2 \tag{5}$$

$$PSNR = 10 \, log\left(\frac{MAX_I^2}{MSE}\right) \tag{6}$$

$$NCC = \frac{\sum_{i=1}^{n}(X_i - \mu_X)(Y_i - \mu_Y)}{\sqrt{\sum_{i=1}^{n}(X_i - \mu_X)^2} \sqrt{\sum_{i=1}^{n}(Y_i - \mu_Y)^2}} \tag{7}$$

## 3. RESULTS

Synthetic T1C MRI generated by the TA-ViT model with both T1W and T2-FLAIR MRI and with T1W only, as well as MRP-ViT, were compared against the ground truth T1C MRI volumes. These were evaluated through the NMSE, PSNR, and NCC metrics (Table 1). In addition, the segmentation performance of the MRP-ViT architecture was measured quantitatively (Table 2). Compared with the MRP-ViT architecture, TA-ViT showed a statistically significant improvement ($p$-value < .001) across all metrics for both the tumor volumes and the entire brain volume, achieving an NMSE: 0.0009 ± 0.0005, PSNR: 31.2 ± 2.2, NCC: 0.908 ± 0.041 for the whole brain. However, there was no statistically significant difference between using multi-parametric inputs and using T1W only. Figure 3 similarly shows the distributions of the tumor and whole brain regions for TA-ViT and MRP-ViT. TA-ViT outperforms MRP-ViT and has fewer very poor reconstructions visualized as a smaller tail for the PSNR and NMSE evaluation metrics.

Example cases of output images from 4 patients are shown in Figure 2. The TA-ViT model produces visually improved tumor regions even for (a) where the predicted segmentation map deviated significantly from the ground truth segmentation map. Table 2 shows the performance of the MRP-ViT model for the T1C-based segmentation task using T1W and T2-FLAIR MRI as input channels.

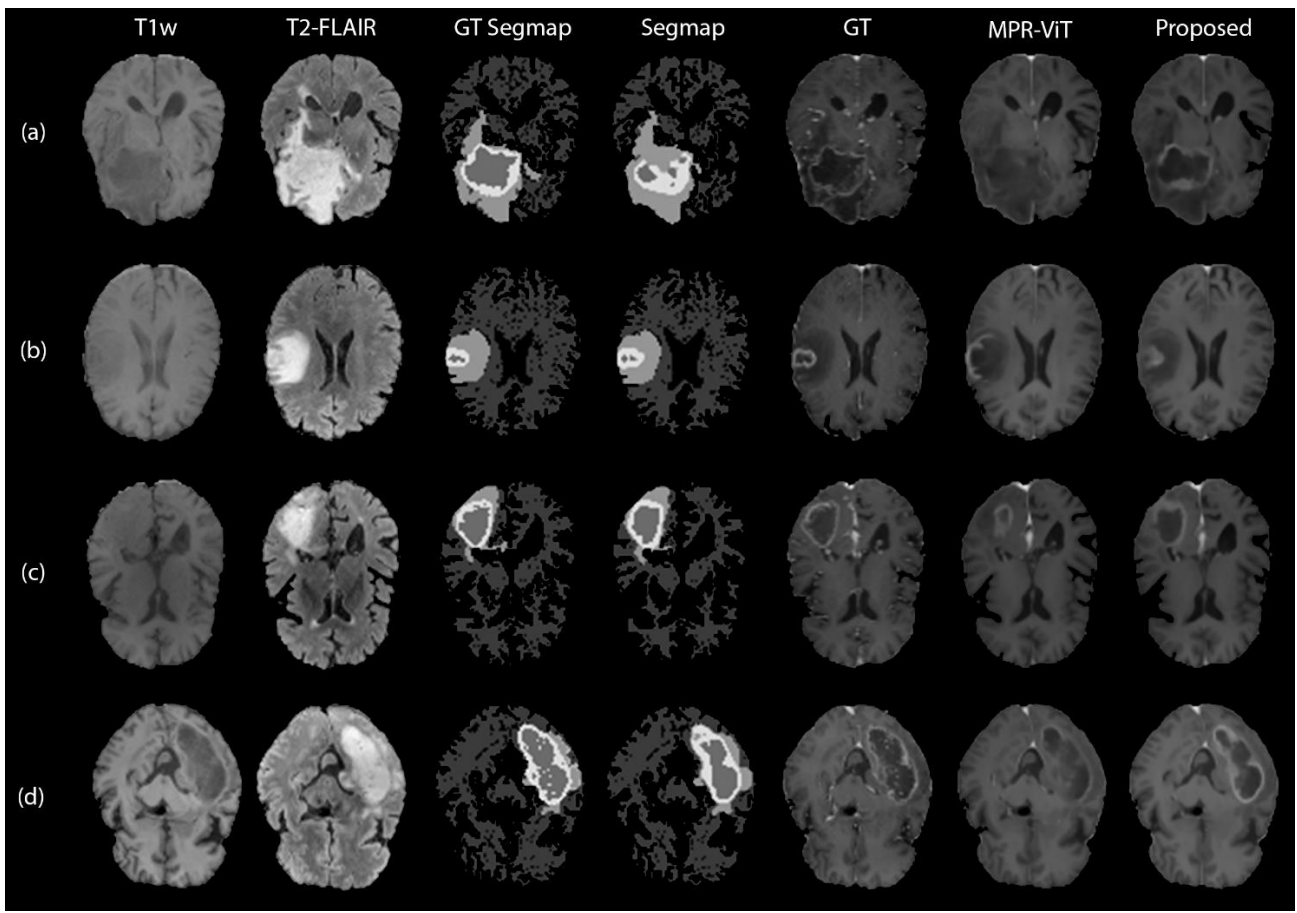

**Figure 2.** Example T1C axial slices generated from ResViT, VCT, and TA-ViT from 4 patients along with zoom-in on interesting regions. All methods took both T1w and T2-FLAIR MRI as inputs. (a) - (d) show reconstructions of slices containing tumor. Shown in the segmentation map in order from darkest to lightest are the necrotic tumor core, brain mask, peritumoral edema, and enhancing tumor.

**Table 1**. Quantitative results for the whole brain volume. P-value metrics are compared against TA-ViT's performance. The arrows indicate the direction of the better metric value. NMSE values are multiplied by $10^{-4}$.

|  | **NMSE[↓] (×10⁻⁴)** |  | **PSNR (dB) [↑]** |  | **NCC[↑]** |  |
|---|---|---|---|---|---|---|
| **Brain** | Mean + Std. | P-value | Mean + Std. | P-value | Mean + Std. | P-value |
| MPR-ViT | 1.21 ± 0.86 | <.001 | 29.9 ± 2.4 | <.001 | .879 ± .042 | <.001 |
| Proposed (T1W) | **8.34 ± 4.85** | 0.261 | **31.4 ± 2.2** | .077 | **908 ± .041** | .150 |
| Proposed (T1W + FLAIR) | 8.53 ± 4.61 | x | 31.2 ± 2.2 | x | **.908 ± .041** | x |
| **Tumor** |  |  |  |  |  |  |
| MPR-ViT | 3.09 ± 2.78 | <.001 | 37.2 ± 5.0 | <.001 | .823 ± .084 | <.001 |
| Proposed (T1W) | **1.20 ± 1.18** | .496 | **41.4 ± 4.8** | .835 | .861 ± .085 | .009 |
| Proposed (T1W + FLAIR) | 1.23 ± 1.28 | x | 41.3 ± 4.7 | x | **.879 ± .042** | x |

**Table 2.** Quantitative results for the tumor region. P-value metrics are compared against TA-ViT's performance. The arrows indicate the direction of the better metric value.

| | DSC[↑] | Jaccard[↑] | RMSD[↓] |
|---|---|---|---|
| MRP-ViT | 0.914 ± .104 | 0.854 ± .135 | 0.031 ± .012 |

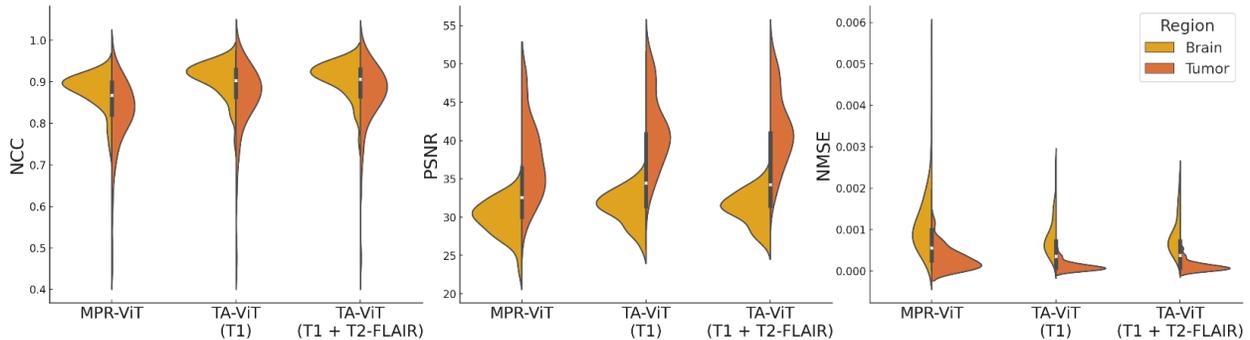

**Figure 3.** Violin plots comparing MPR-ViT, TA-ViT (T1W), and TA-ViT (T1W + T2-FLAIR) for NCC, PSNR, and NMSE for both the tumor and whole brain regions

## 4. DISCUSSION

In this study, we propose the TA-ViT model for multimodal T1C synthesis which outperforms the state-of-the-art MPR-ViT model as measured by the NMSE, PSNR, and NCC and based on qualitative results. By generating accurate T1C MRI from mp-MRI, potential toxicity from GBCAs can be prevented and may be a substitute when T1C MRI is not practical to obtain. The synthetic T1C MRI images generated by the TA-ViT model are, overall, highly conformal to the ground truth tumor and brain volumes even in difficult heterogeneous tumor regions. These favorable image characteristics may be useful for tumor characterization and detection and motivate further investigation.

To our knowledge, this is the first work to synthesize T1C MRI from non-contrast MRI with a focus on tumor region reconstruction. However, several notable studies have been published for multi-modal and T1C MR image translation tasks.[31] Liu et al. proposed the multi-contrast multi-scale Transformer (MMT) comprised only of Swin-Transformer blocks and achieved their best results using T1W, T2W, and T2-FLAIR MRI to predict T1C, achieving a PSNR, SSIM, and learned perceptual image patch similarity (LPIPS) score of 29.74, .939, and .120 respectively.[32,33] The synthetic T1C images were then segmented using the top-performing algorithm in the BraTS 2021 challenge and obtained a DSC of .726 ± .137 for the whole tumor (WT) volume.[34] In addition, Osman and Tamam proposed a 3D dense-dilated residual U-Net (DD-Res U-Net to synthesize T1C MRI from T1W, T2W, and T2-FLAIR MRI, achieving a PSNR 30.284 ± 4.934, a SSIM of 0.915 ± 0.063, and a MSE of 0.001 ± 0.002 for the whole brain region. However, no tumor information was given to the network during training, so performance on the tumor region was found to be suboptimal.

Comparing the input T1w and T2-FLAIR MRI with the synthetic T1s, this study reveals that the synthetic T1C images have superior contrast of the tumor region, especially in delineating the tumor boundary. In addition, the TA-ViT model demonstrated the ability to recover detail only visible in ground truth T1C such as in Figure 2b. However, we note that the T1C reconstructions are imperfect, especially in heterogeneous regions. This is best shown in Figure 2A where small amounts of contrast were not captured outside of the enhancing tumor region. An interesting result is that the NCC values for all model predictions were better for the whole brain volume compared to the tumor region while the NMSE and PSNR values were better for the tumor volume. We speculate that this is because the NCC looks for similarities in structure, which can be irregular or lacking in the tumor regions, whereas the relative homogeneity of the tumor region may benefit MSE and PSNR. In contrast with previous work showing improved performance with multi-modal inputs,[35] there was no significant difference between inputting only T1W MRI and both T1W MRI and T2-FLAIR. This would be advantageous for predicting synthetic T1C MRI in a clinical setting since the workflow is simplified and there is no need for registration between T1W and T2-FLAIR MRI.

We acknowledge several limitations of the present work. The TA-VIT model presented here is trained on 2D slices and so does not directly capture the full 3D context of the input data. While the convolutional layers employed in TA-VIT provide an efficient way to capture the fine details, they might also miss key details due to their short-range context. Future work may involve replacing the encoder and decoder by performing the training in a latent space representation as was done with the segmentation maps in this study and removing the convolutional layers in the information bottleneck to create a ViT-only architecture. The dataset was comprised of only glioma patients, so it remains to be seen how the results will generalize to patients with other diseases. However, given the successful application of related image translation tasks in these settings, we are optimistic about the generalizability of this model.[36-38] We intend in future work to incorporate full 3D context, train on larger more diverse datasets as they become available[22,39], explore diffusion models[40], and see if synthetic T1C can be used to differentiate LGG from HGG. In addition, we would also like to condition the model on genomics or clinical data in addition to segmentation maps to see if they further improve model performance.

## 5. CONCLUSION

This study presents a deep hybrid CNN-transformer model designed to accurately predict glioma tumor volumes for multimodal T1C synthesis. By conditioning the ViT blocks on predicted segmentation maps with the adaLN-zero mechanism, the reconstructive ability of the network was significantly increased with minimal changes in computational complexity. The proposed method shows great promise in making valuable information in T1C MRI more readily available and may serve as an example for other image synthesis tasks interested in accurate tumor region reconstruction.

## ACKNOWLEDGMENTS

This research is supported in part by the National Cancer Institute of the National Institutes of Health under Award Numbers R01CA272991, R56EB033332, R01DE33512 and R01CA272755. This work was supported in part by Oracle Cloud credits and related resources provided by Oracle for Research.

## Disclosures

The author declares no conflicts of interest.